\batchmode
\documentclass[twoside,fleqn,espcrc2]{article}
\usepackage{espcrc2}
\documentstyle[twoside,fleqn,espcrc2]{article}

\newcommand{\beq}{\begin{equation}}
\newcommand{\ber}{\begin{eqnarray}}
\newcommand{\eeq}{\end{equation}}
\newcommand{\eer}{\end{eqnarray}}

\newcommand{\AmS}{{\protect\the\textfont2
  A\kern-.1667em\lower.5ex\hbox{M}\kern-.125emS}}

\title{Tau as a Probe for New Physics  \hfill  SUNY BING
10/15/94}

\author{Charles A. Nelson\address{Department of Physics, State
University of
New York at Binghamton,\\
	Binghamton, New York 13902-6000}%
      \thanks{cnelson@bingvmb.cc.binghamton.edu \newline
                   3rd Tau Workshop, Sept. '94,Montreux} }

\begin{document}

\begin{abstract}
The usage of polarimetry and spin-correlation tests to determine the
{\em complete Lorentz structure} of the tau lepton's charged and
neutral-current couplings is reviewed.  The emphasis is on tests for
``something'' in a $(V-A) + $ ``something'' structure in
${J^{Charged}}_{Lepton}$ current, so as to bound the scales
$\Lambda$ for ``new physics" such as arising from tau weak
magnetism,  weak electricity, and/or second-class currents.  Tests for
$T$ and for $CP$ violation are discussed.
\end{abstract}

\maketitle

\section{TESTS FOR \newline  \mbox{``ADDITIONAL
STRUCTURE''} TO
\newline THAT OF THE \mbox{STANDARD MODEL} }

There are many exciting directions in which the tau leptons,
$\tau^\pm, \nu_\tau,
\={\nu}_\tau$ , can be used to probe for new physics beyond that in
the standard
model.  Here we will focus on and review the usage of polarimetry
and spin-
correlation tests \cite{a1,b1,mc1,a2,c1,a4}  to determine the {\em
complete
Lorentz structure} of the tau lepton's charged and neutral-current
couplings.  Tests
for time-reversal violation and for CP violation are thereby included.

For charged-current couplings, the emphasis will be on
\cite{a3,C92,C94,C94a}
\begin{itemize}
    \item Tests for ``something'' in a { \begin{quote}
(V-A) + ``something''
\end{quote} } structure in the ${J^{Charged}}_{Lepton}$ current.
    \item Tests for a non-CKM-type leptonic CP \newline violation in
$ \tau
\rightarrow \rho \nu $ and $ \tau \rightarrow a_{1} \nu $ decays.
\end{itemize}

For neutral-current couplings, the emphasis will be on
\begin{itemize}
    \item Tests for ``additional structure'' in both the weak and the
pure
electromagnetic ${J^{Neutral}}_{Lepton}$ currents.
    \item Complete tests for possible CP and/or T violations in
$Z^{o}$, or
$\gamma^{*}\rightarrow \tau^{-} \tau^{+}$.
\end{itemize}
Our discussion of neutral-current couplings will be brief for they
have already
been discussed at this conference by W. Marciano \cite{t1} and A.
Stahl \cite{t2}.

In searching for new physics, it is very important to carefully and
systematically
search for {\em symmetry violations}, such as an unexpected
Lorentz structure in
tau couplings,  a leptonic CP or T violation, a finite $\nu_\tau$ mass,
a violation of
``lepton number'' conservation or of $L_\tau , L_\mu$, or $L_e$
conservation.
Such violations may show up by the observation of an otherwise
forbidden decay
mode, or though polarimetry and spin-correlation tests.

A healthy ``working attitude'' is to remember that we will only see
the {\em
symmetry violations} which the observables allow.  Hence, if a
violation exists,
until we uncover the right observable it won't be seen.  If it's a small
effect (for
some reason), it still won't be seen.  However, {\em if it isn't
fundamentally
forbidden, it will make itself seen somewhere, sometime}!

\section{``ADDITIONAL STRUCTURE'' IN \newline
CHARGED-CURRENT COUPLINGS?}

\subsection{The simplest S2SC function}

\begin{figure*}[t]
\vspace{8cm}
\caption{Illustration of the two stage structure of the stage-two spin-
correlation
(S2SC) function. $ \rho$ polarimetry information from the next stage
of decays
$\rho^{ch} \rightarrow \pi^{ch} \pi^{0}$ is included in the helicity
amplitude
treatment of the production-decay sequence $e^{-} e^{+}
\rightarrow Z^{0},
\gamma^{*} \rightarrow \tau_{1}^{-} \tau_{2}^{+} \rightarrow
(\rho^{-} \nu)
(\rho^{+} \=\nu)$. Compare Fig. 3.}
\label{fig:s2sc}
\end{figure*}

The idea is to search for ``additional structure'' in
${J^{Charged}}_{Lepton}$ by
generalizing the $\tau $ spin-correlation function $I(E_\rho ,E_B)$
by
including the $\rho $ polarimetry information \cite{a2} that is
available from the
$%
\rho^{ch}\rightarrow \pi^{ch}\pi ^o$decay distribution
\cite{C92,C94}. The
symbol $B=\rho
,\pi ,l$ . Since this adds on spin-correlation information from the
next
stage of decays in the decay sequence, we call such an energy-
angular
distribution a stage-two spin-correlation (S2SC) function. Compare
Fig.1.

The simplest useful S2SC is for the $CP$-symmetric decay sequence
$Z^o$, or
$%
\gamma ^{*}\rightarrow \tau ^{-}\tau ^{+}\rightarrow (\rho ^{-}\nu
_\tau
)(\rho ^{+}\bar \nu _\tau )$ followed by both $\rho ^{\mp
}\rightarrow \pi
^{\mp }\pi ^o$.

$$
\begin{array}{lcc}
\text{I(E}_\rho \text{,E}_{\bar \rho }\text{,}\tilde \theta
_1\text{,}\tilde
\theta _{2\text{ }}\text{) = }|\text{T}\left( +-\right) |^2\rho
_{++}\bar \rho _{--} \\
+\text{ }|\text{T}\left( -+\right) |^2\rho _{--}\bar \rho
_{++}+\text{ }|\text{T}\left( ++\right) |^2\rho _{++}\bar \rho _{++}
 \\+\text{ }|\text{T}\left( --\right) |^2\rho
_{--}\bar \rho _{--}
\end{array}
$$
If we think in terms of probabilities, the quantum-mechanical
structure of this
expression for $I_4$ is almost intuitively obvious,  since the
$T(\lambda _{\tau
^{-}},\lambda _{\tau ^{+}})$ helicity amplitudes
describe the production of the $(\tau ^{-}\tau ^{+})$ pair via
$Z^o$or $%
\gamma ^{*}\rightarrow \tau ^{-}\tau ^{+}$. So for instance in the
1st term,
the factor  $|T(+,-)|^2=$``Probability to produce a $\tau ^{-}$ with
$%
\lambda _{\tau ^{-}}=\frac 12$ and a $\tau ^{+}$with $\lambda
_{\tau
^{+}}=-\frac 12$ '' is simply multiplied by the product of the decay
probablity, $\rho _{++}$, for the positive helicity $\tau ^{-
}\rightarrow
\rho ^{-}\nu \rightarrow \left( \pi ^{-}\pi ^o\right) \nu $ times the
decay
probablity, $\bar \rho _{--}$, for the negative helicity $\tau
^{+}\rightarrow \rho ^{+}\bar \nu \rightarrow \left( \pi ^{+}\pi
^o\right)
\bar \nu $ .

\begin{figure*}[t]
\vspace{5.5cm}
\caption{For ``stage 1'', the three angles $\theta_1^\tau,
\theta_2^\tau$ , and $\phi$
describe the sequential decay $Z^{0}$ or $\gamma^{*} \rightarrow
\tau_{1}^{-}
\tau_{2}^{+}$ with $\tau_{1}^{-} \rightarrow \rho_{1}^{-} \nu$
and
$\tau_{2}^{+} \rightarrow \rho_{2}^{+} \=\nu$ .  From (a) a boost
along the
negative $z_{1}^{\tau}$ axis transforms the kinematics from the
$\tau_{1}^{-}$
rest frame to the $ Z^{0} / \gamma^{*}$ rest frame and, if boosted
further, to the
$\tau_{2}^{+}$ rest frame in (b).}
\label{fig:var1}
\end{figure*}

What are the kinematic variables? The 1st stage of the decay
sequence $\tau
^{-}\tau ^{+}\rightarrow (\rho ^{-}\nu _\tau )(\rho ^{+}\bar \nu
_\tau )$ is
described by the 3 variables%

$$
\begin{array}{ccc}
\theta _1^\tau  & \equiv  & \text{E}_\rho  \\ \theta _2^\tau  & \equiv
&
\text{E}_{\bar \rho } \\ \cos \phi  & \equiv  & \cos \psi
\end{array}
$$
where the helicity variables $\theta _1^\tau ,\theta _2^\tau ,\cos \phi
$
are defined as in Fig.2.

 These are equivalent to the $Z^o$, or $\gamma ^{*}$
center-of-mass variables, $E_{\rho}, E_{\bar \rho }, \cos \psi $.
Here $\psi =$%
``opening $\angle $ between the $\rho ^{-}$ and $\rho
^{+}$momenta in the $%
Z/\gamma ^{*}$ cm''.

When the Lorentz ``boost'' to one of the $\rho $ rest frames is
directly from the
$Z/\gamma
^{*}$ cm frame, the 2nd stage of the decay sequence is described by
the
usual 2 spherical angles, see Fig.3, for the $\pi ^{ch}$ momentum
direction in that $\rho $ rest frame:
\newpage

For $\rho _1^{-}\rightarrow \pi _1^{-}\pi _1^o$: \hspace{2pc} For
$\rho
_2^{+}\rightarrow \pi _2^{+}\pi _2^o$: \\
 \hspace*{3pc} $\tilde \theta _1, \tilde \phi _1$ \hspace{7pc} $\tilde
\theta _2,
\tilde \phi _2$
\\

What are the dynamical variables? More precisely the $T(\lambda
_{\tau
^{-}},\lambda _{\tau ^{+}})$ helicity amplitudes describing the
production
of the $(\tau ^{-}\tau ^{+})$ pair via $Z^o$or $\gamma
^{*}\rightarrow \tau
^{-}\tau ^{+}$, are defined by the associated matrix element%

$$
\langle \Theta _\tau ,\Phi _\tau |\text{J=1},\text{M}\rangle
=\text{D}%
_{M\lambda }^{1*}(\Phi _\tau ,\Theta _\tau ,-\Phi _\tau )\text{
T}\left(
\lambda _1,\lambda _2\right)
$$

\begin{figure*}[t]
\vspace{11.2cm}
\caption{For ``stage 2'', the spherical angles $\~{\theta}_{1},
\~{\phi}_{1}$
specify the $\pi^{-}$ momentum in $\rho_{1}^{-} \rightarrow \pi^{-
} \pi^{0}$
decay in the $\rho_{1}^{-}$ rest frame when the boost is directly
from the
$Z^{0}$ or $\gamma^{*}$ cm frame.  Similarly, $\~{\theta}_{2},
\~{\phi}_{2}$
specify the $\pi^{+}$ momentum in $\rho_{2}^{+} \rightarrow
\pi^{+} \pi^{0}$.
The $\rho^{-} \rho^{+}$ production half-plane specifies the positive
$x_1$ and
$x_2$ axes.}
\label{fig:var2}
\end{figure*}

The composite decay density matrix elements are given by ( for only
$\nu _L$
couplings)%

$$
\begin{array}{lllll}
\rho _{hh}=
 \\ \left( 1+h\cos \theta _1^\tau \right) \left[ \cos ^2\omega _1\cos
^2\tilde \theta _1+\frac 12\sin ^2\omega _1\sin ^2\tilde \theta
_1\right]
\\
+
\frac{r_a^2}2\left( 1-h\cos \theta _1^\tau \right) \left[ \sin ^2\omega
_1\cos ^2\tilde \theta _1  \\ +\frac 12\left( 1+\cos ^2\omega _1\right)
\sin
^2\tilde \theta _1 ]  \\ +h\frac{r_a}{\sqrt{2}}\cos \beta _a\sin \theta
_1^\tau \sin 2\omega _1\left[ \cos ^2\tilde \theta _1-\frac 12\sin
^2\tilde
\theta _1\right]
\end{array}
$$
They are, however, simply the decay probability for a $\tau _1^{-}$
with
helicity $\frac h2$ to decay $\tau ^{-}\rightarrow \rho ^{-}\nu
\rightarrow
\left( \pi ^{-}\pi ^o\right) \nu $ since

\beq
\frac{d\text{N}}{d\left( \cos \theta _1^\tau \right) d\left( \cos \tilde
\theta _1\right) }=\rho _{hh}\left( \theta _1^\tau ,\tilde \theta
_1\right)
\eeq

For the $CP$-conjugate decay of the $\tau _2^{+}$ ,

\beq
\bar \rho _{hh}=\rho _{-h,-h}\left( \text{subscripts} 1 \rightarrow 2,
a \rightarrow
b \right)
\eeq

It is more fun to discuss the remaining quantities ($\beta_a, r_a,
\omega_1 $) in
$\rho _{hh\text{ }}$ in
answer to the question ``What is the pure $\nu _L$'s  $I(E_\rho
,E_{\bar
\rho },\tilde \theta _1,\tilde \theta _2)$ good for?  The answer is that
one
obtains 2 tests for ``non-CKM-type leptonic $CP$ violation''.  This is
explained in
the next section.

However, there is even a simpler interesting application of $I_4$:  At
this
conference, there has been considerable discussion about different
methods for the
determination of the mass of
the $\tau $ neutrino. The full $I_4$ spin-correlation function
including $\nu_R$
and $\nu_L$ couplings (see Sec. 2.3) depends on
the ratio $\left( \frac{m_\nu }{m_\tau }\right) ^2$. So before using
any $I_4$ for
other purposes where we will set $m_\nu = 0$,  it is important to
check the
sensitivity to $m_\nu
\neq 0$. The present experimental limit $m_\nu = 24MeV$
corresponds to
\beq
\delta \left( (\frac{m_\nu }{m_\tau })^2\right) =2\frac{m_\nu
}{m_\tau }%
\delta (\frac{m_\nu }{m_\tau })=6\cdot 10^{-4}
\eeq
Table 1 gives the sensitivity of the full $I_4$ to  $\left( \frac{m_\nu
}{m_\tau }%
\right) ^2$.

\begin{table}[hbt]
% space before first and after last column: 1.5pc
% space between columns: 3.0pc (twice the above)
\setlength{\tabcolsep}{.3pc}
% -----------------------------------------------------
% adapted from TeX book, p. 241
\newlength{\digitwidth} \settowidth{\digitwidth}{\rm 0}
\catcode`?=\active \def?{\kern\digitwidth}
% -----------------------------------------------------
\caption{Sensitivity of $I(E_1,E_2,\tilde{\theta_1},\tilde{\theta_2})$
to
$m_{\nu}\neq 0$:}
\label{tab1}
\begin{tabular}{lrr}
\hline
\\ [.5pt]
                                                                  &
\multicolumn{2}{l}{Pure $V-A$
coupling} \\
\cline{2-3}
                                                                  &
\multicolumn{1}{r}{At
$M_Z$}
                                                                  &
\multicolumn{1}{r}{10, or 4 GeV}
\\
\hline
{      $\lbrace \rho^{-}, \rho^{+} \rbrace$ mode:}      &
&
\\

&$70\cdot
10^{-
4}$&$13\cdot 10^{-4}$   \\
{      $\lbrace a_{1}^{-}, a_{1}^{+} \rbrace$ mode:}&
&
\\

&$18\cdot
10^{-4}$&$
3\cdot 10^{-4}$   \\
\hline
%\multicolumn{3}{@{}p{20mm}}{}
\end{tabular}
\end{table}
All tables in this paper list only the ideal statistical errors \cite{c1},
and assume
respectively $10^7 Z^{o}$events and $10^7$ (\tau^{-} \tau^{+}
)$pairs.

At $10GeV$, or $4GeV$ the numbers correspond to $613,387$
events for $\{\rho
^{-},\rho ^{+}\}$or B($\tau \rightarrow \rho \nu $) = $24.6$\% , and
$324,000$
events for $\{a_1^{-},a_1^{+}\}$.
Throughout this paper for the $a_1$ mode we sum the charged plus
neutral
pion $a_1$ final states so B($\tau \rightarrow {a_1}^{ch+neu} \nu
$) = $18$\%,
and use $m_{a_1} = 1.275GeV$.  Note that Table 1 shows the
expected
substantial improvement in the
limit on $m_\nu $ upon using the greater $a_1$ mass, instead of the
$\rho$ mass.
When
systematic errors are included, the numbers shown in  Table 1 mean
that
multi-pion modes must be used. For the analysis in Table 1, we did
not need
form-factor assumptions based on current algebra type arguments;
perhaps
eventually
form-factor independent measurements will be performed for the
multi-pion $%
\tau $ decay modes.

\subsection{Tests for non-CKM-type leptonic CP violation}

It is simple to see from the kinematics why two $CP$ tests are
possible:  First we
again start with the classic definitions for the necessary helicity
amplitudes in each
decaying particle's rest frame.  In the  $\tau ^{-}$ rest frame, the
matrix element
for $\tau ^{-}\rightarrow \rho ^{-}\nu$ is

$$
\langle \theta _1^\tau ,\phi _1^\tau ,\lambda _\rho ,\lambda _\nu
|\frac
12,\lambda _1\rangle =D_{\lambda _1,\mu }^{\frac 12*}(\phi
_1^\tau ,\theta
_1^\tau ,0)A\left( \lambda _\rho ,\lambda _\nu \right)
$$
where $\mu =\lambda _\rho -\lambda _\nu $.  Similarly, in the $\rho
^{-}$ rest
frame

\beq
\langle \tilde \theta _a,\tilde \phi _a|\lambda _\rho \rangle
=D_{\lambda
_\rho,0}^1(\tilde \phi _a,\tilde \theta _a,0)c
\eeq
where $c$ is a constant factor.  And, for the $CP$-conjugate
process, $\tau
^{+}\rightarrow \rho ^{+}\bar \nu \rightarrow \left( \pi ^{+}\pi
^o\right) \bar \nu $, in the $\tau ^{+}$ rest frame
$$
\langle \theta _2^\tau ,\phi _2^\tau ,\lambda _{\bar \rho },\lambda
_{\bar
\nu }|\frac 12,\lambda _2\rangle =D_{\lambda _2,\bar \mu }^{\frac
12*}(\phi
_2^\tau ,\theta _2^\tau ,0)B\left( \lambda _{\bar \rho },\lambda
_{\bar \nu
}\right)
$$
with $\bar \mu =\lambda _{\bar \rho }-\lambda _{\bar \nu }$.  In the
$\rho ^{+}$
rest frame

\beq
\langle \tilde \theta _b,\tilde \phi _b|\lambda _{\bar \rho }\rangle
=D_{\lambda _{\bar \rho,0 }}^1(\tilde \phi _b,\tilde \theta _b,0)\bar
c
\eeq

Assuming a L-handed $\nu _\tau $, $\tau ^{-}\rightarrow \rho ^{-
}\nu $
depends on
$$
\begin{array}{c}
A\left( -1,-\frac 12\right) =|A\left( -1,-\frac 12\right) |
\text{ e}^{\iota \phi _{-1}^a} \\ A\left( 0,-\frac 12\right) =|A\left(
0,-\frac 12\right) |
\text{ e}^{\iota \phi _0^a}
\end{array}
$$
and $ A\left( 1,-\frac 12\right) =0$ by rotational invariance.  In this
section we will
neglect the R-handed $\nu _\tau $ amplitudes. Note
that the R-handed $\nu _\tau $ amplitudes for pure $(V-A)$ are
of order
$$
\begin{array}{c}
\frac{A\left( 1,\frac 12\right) }{A\left( -1,-\frac 12\right) }\simeq
\frac{%
m_\nu m_\tau }{m_\tau ^2-m_\rho ^2} \\ \frac{A\left( 0,\frac
12\right) }{%
A\left( 0,-\frac 12\right) }\simeq \frac{m_\nu m_\rho ^2}{m_\tau
(m_\tau
^2-m_\rho ^2)}
\end{array}
$$

Similarly for the CP-conjugate process, assuming a R-handed $\bar
\nu _\tau $%
, $\tau ^{+}\rightarrow \rho ^{+}\bar \nu $ depends on

$$
\begin{array}{c}
B\left( 1,\frac 12\right) =|B\left( 1,\frac 12\right) |
\text{ e}^{\iota \phi _1^b} \\ B\left( 0,\frac 12\right) =|B\left( 0,\frac
12\right) |
\text{ e}^{\iota \phi _0^b} \end{array}
$$
and $B\left( -1,\frac 12\right) =0$.

\begin{table*}[t]
% space before first and after last column: 1.5pc
% space between columns: 3.0pc (twice the above)
\setlength{\tabcolsep}{1.5pc}
% -----------------------------------------------------
% adapted from TeX book, p. 241
\newlength{\digitwidth} \settowidth{\digitwidth}{\rm 0}
\catcode`?=\active \def?{\kern\digitwidth}
% -----------------------------------------------------
\caption{2 Tests for ``Non-CKM-Type Leptonic CP Violation'' in
$\tau \rightarrow
\rho \nu $ decay (ideal statistical errors):}
\label{tabcpr}
\begin{tabular*}{\textwidth}{@{}l@{\extracolsep{\fill}}rrr}
\hline
\\ [.5pt]
$E_{cm}$  & $\sigma (r_a)$                   &$\sigma (\~{\beta})\simeq
\sigma
(\beta_a)$ & $\sigma ( \beta^{\prime})$ \\
                    &{\bf $CP\~{T_{FS}},CP$}&{\bf
$CP$\hspace{2pc}$\~{T_{FS}$
}         &{\bf $CP\~{T_{FS}},CP$} \\
\hline
$M_Z$       & $0.6\% $                             & $\sim 1.9^{o}$
&  $\sim 3^{o}$ \\
$10 GeV$  & $0.1\% $                             & $\sim 0.4^{o}$
& $\sim 0.7^{o}$ \\
$4 GeV$    & $0.1\% $                             & $\sim 0.9^{o}$
& $\sim 1.1^{o}$ \\
\hline
\\ [.5pt]
\multicolumn{4}{@{}p{120mm}}{Note $\~{\beta} = \beta_a -
\beta_b$ and
$\beta^{\prime} = \beta_a + \beta_b$.}
\end{tabular*}
\end{table*}

By CP invariance, the 2 tests for non-CKM-type leptonic CP
violation are: %
$$
\beta _a=\beta _b \hspace{2pc} {\bf first \hspace*{.4pc} test}
$$
where $\beta _a=\phi _{-1}^a-\phi _0^a$, $\beta _b=\phi _1^b-\phi
_0^b$,and
$$
r_a=r_b \hspace{2pc} {\bf second  \hspace*{.4pc} test}
$$
where%
$$
r_a=\frac{|A\left( -1,-\frac 12\right) |}{|A\left( 0,-\frac 12\right)
|},r_b=%
\frac{|B\left( 1,\frac 12\right) |}{|B\left( 0,\frac 12\right) |}
$$
This easily follows since in the helicity formalism the symmetries
for  $%
\tau ^{-}\rightarrow \rho ^{-}\nu $ and $\tau ^{+}\rightarrow \rho
^{+}\bar
\nu $ are:%
$$
\begin{array}{cc}
\underline{Invariance} & \underline{Relation} \\ P & A\left( -
\lambda _\rho
,-\lambda _\nu \right) =A\left( \lambda _\rho ,\lambda _\nu \right)  \\
& B\left( -\lambda _{\bar \rho },-\lambda _{\bar \nu }\right) =B\left(
\lambda _{\bar \rho },\lambda _{\bar \nu }\right)  \\
C & B\left( \lambda _{\bar \rho },\lambda _{\bar \nu }\right)
=A\left(
\lambda _{\bar \rho },\lambda _{\bar \nu }\right)  \\
CP & B\left( \lambda _{\bar \rho },\lambda _{\bar \nu }\right)
=A\left(
-\lambda _{\bar \rho },-\lambda _{\bar \nu }\right)  \\
\tilde T_{FS} & A^{*}\left( \lambda _\rho ,\lambda _\nu \right)
=A\left(
\lambda _\rho ,\lambda _\nu \right)  \\
& B^{*}\left( \lambda _{\bar \rho },\lambda _{\bar \nu }\right)
=B\left(
\lambda _{\bar \rho },\lambda _{\bar \nu }\right)  \\
CP\tilde T_{FS} & B^{*}\left( \lambda _{\bar \rho },\lambda _{\bar
\nu
}\right) =A\left( -\lambda _{\bar \rho },-\lambda _{\bar \nu }\right)
\end{array}
$$
There is a basic theorem that measurement of a non-real helicity
amplitude implies
a violation of $\tilde T_{FS}$ invariance when a first-order
perturbation in an
``effective" hermitian Hamiltonian is reliable.  So $\tilde T_{FS}$
invariance is
expected to be violated when there are significant final-state
interactions; and it is
to be distinguished from canonical $T$ invariance which requires
interchanging
``final'' and ``initial'' states, i.e. actual time-reversed reactions are
required.

In the standard lepton model with a pure $(V-A)$ coupling, the
values of the
these polar parameters are $\beta _a=0,r_a=\frac{\sqrt{2}m_\rho
}{E_\rho
+q_\rho }\simeq \sqrt{2}m_\rho /m_\tau \simeq 0.613.$

These 2 tests should be compared with the classic CP test for the
equality
of the partial widths of CP-conjugate reactions $A_\Gamma \equiv
\frac{%
\Gamma -\bar \Gamma }{\Gamma +\bar \Gamma }$. Where, e.g. for
$\tau
\rightarrow \rho \nu $ decay, $\Gamma =\Gamma \left( \tau ^{-
}\rightarrow
\rho ^{-}\nu \right) $ and  $\bar \Gamma =\bar \Gamma \left( \tau
^{+}\rightarrow \rho ^{+}\bar \nu \right) $.  So, $A_\Gamma $ tests
for
CKM-type CP violation and for $r_a/r_b\neq 1$, but it is
\underline{not}
sensitive to $\beta _a\neq \beta _b$.

In contrast, $I_4$'s two tests for $%
r_a/r_b\neq 1$ and  $\beta _a\neq \beta _b$ are tests for a
non-CKM-type of leptonic CP violation.  Any overall leptonic
CKM-type
phases in any mixture of V and A couplings will equally effect the
$A\left(
-1,-\frac 12\right) $and  $A\left( 0,-\frac 12\right) $amplitudes and so
will cancel out in $r_a$ and in $\beta _a$. (However, for $S,P,T$
and $T_5$
couplings,
CKM-phases are in general observable by S2SC functions.)
Quantitatively, for $A_{amp}=A+\delta a,\bar A_{amp}=A$, then
$A_\Gamma
=\delta a/A\geq \left( 1-4\%\right) $ since for $\tau $ 2-body modes
the
denominator is known to $1-4\%$. Whereas, as shown in Table 2 for
$%
\tau ^{-}\rightarrow \rho ^{-}\nu $ , $\delta r_a/r_a=\left( 0.1-
1\%\right) $.

\begin{table}[hbt]
% space before first and after last column: 1.5pc
% space between columns: 3.0pc (twice the above)
\setlength{\tabcolsep}{.4pc}
% -----------------------------------------------------
% adapted from TeX book, p. 241
\newlength{\digitwidth} \settowidth{\digitwidth}{\rm 0}
\catcode`?=\active \def?{\kern\digitwidth}
% -----------------------------------------------------
\caption{Comparison of discrete symmetry requirements for moduli
ratios and
phase differences:}
\label{tabs}
\begin{tabular}{lrr}
\hline
\\ [.5pt]

\multicolumn{1}{c}{\bf
$CP\~{T_{FS}}$}
                                                                  &
\multicolumn{1}{c}{\bf $CP$ }
                                                                  &
\multicolumn{1}{c}{\bf
$\~{T_{FS}}$}    \\
\hline
$r_a = r_b $                                                        &    $r_a =
r_b $
&  No
prediction   \\
                                                                             &
{\bf same}
&
\\
$\beta_a = - \beta_b$,                                        &  $\beta_a =
\beta_b$  &  All
$\beta$'s vanish  \\
\hspace{.5pc} or $\beta_a + \beta_b = 2 \pi$  &          {\bf opposite}
&        \\
\hline
\end{tabular}
\end{table}

Table 3 shows that the $CP$ and $CP\tilde T_{FS}$ predictions for
the
phase relation between $\beta _a$ and $\beta _b$ are opposite. So
this
provides a method for distinguishing between a new physics effect
due to an
unusual $CP$-violating final state interaction and one with a
different
mechanism of $CP$ violation.

In $\rho_{hh}$, the $\omega _1$ parameter is only a function of
$\tilde \theta _1$
(i.e. of $%
E_\rho $) since%
$$
\sin \omega _1=m_\rho \beta \gamma \sin \theta _1^\tau /{p_1}
$$
$$
\cos \omega _1=\frac{M}{4m_\tau ^2p_1} \left( m_\tau ^2-m_\rho
^2 +\left[
m_\tau ^2+m_\rho ^2\right]
\beta \cos \theta _1^\tau \right)
$$
where $M=E_{cm}$, $\gamma =M/(2m_\tau )$. Physically $\omega
_{1\text{
}}$
characterizes the Wigner rotation angle which occurs in going from
the $\tau
^{-}$ rest frame's description of $\rho ^{-}$ decay ( which is most
easily
used in the helicity formalism) and the $Z^o/\gamma ^{*}$rest
frame's
variables for  $\rho ^{-}$ decay which are most easily measured
experimentally. In the future, with silicon-vertex-detectors the  $\tau
^{-}$
rest frame may be known experimentally for a large sub-sample of
$\tau
^{-}\tau ^{+}$pair events in which case a simpler 4 variable
function $%
I\left( E_{\rho},E_{\bar \rho },\tilde \phi_a,\tilde \phi_b\right) $%
could be used, see ref. \cite{C94}.

\begin{table*}[t]
% space before first and after last column: 1.5pc
% space between columns: 3.0pc (twice the above)
\setlength{\tabcolsep}{1.5pc}
% -----------------------------------------------------
% adapted from TeX book, p. 241
\newlength{\digitwidth} \settowidth{\digitwidth}{\rm 0}
\catcode`?=\active \def?{\kern\digitwidth}
% -----------------------------------------------------
\caption{2 Tests for ``Non-CKM-Type Leptonic CP Violation'' in
$\tau^{-}
\rightarrow a^{-}_{1} \nu $ decay, both $(2 \pi^{-} \pi^{+} )$ and
$(2 \pi^{o}
\pi^{-} )$ (ideal statistical errors):}
\label{tabcpa}
\begin{tabular*}{\textwidth}{@{}l@{\extracolsep{\fill}}rrr}
\hline
\\ [.5pt]
$E_{cm}$ & $\sigma (r_a)$                    &$\sigma (\~{\beta})\simeq
\sigma
(\beta_a)$&\sigma (\beta^{\prime})$\\
                    &{\bf $CP\~{T_{FS}},CP$}&{\bf
$CP$\hspace{2pc}$\~{T_{FS}$
}        &{\bf $CP\~{T_{FS}},CP$}\\
\hline
$M_Z$       & $0.3\% $                             & $\sim 10^{o}$
&  $\sim 15^{o}$ \\
$10 GeV$  & $0.05\% $                           & $\sim 3^{o}$
& $\sim  3^{o}$ \\
$4 GeV$    & $0.05\% $                           & $\sim 4^{o}$
& $\sim  5^{o}$ \\
\hline
\end{tabular*}
\end{table*}

Notice that measurement of $\beta _a$ by the simple 4-variable
SCSC function
is possbile only because of the occurence of the Wigner rotation. In
$\rho
_{hh}$, $\cos \beta _a$ appears multiplied by $\cos \omega _1$.
Also notice the
necessity of the $\tau ^{-}$ being polarized for measurement of
$\beta _a$. Here
this is achieved by using the spin-correlation technique. At $\gamma
^{*}$
energies, $\beta _a$ cannot be measured by only analyzing the decay
of an
unpolarized $\tau ^{-}$ via $\tau ^{-}\rightarrow \rho ^{-}\nu
\rightarrow \left( \pi ^{-}\pi ^o\right) $.  At the $Z^o,$ the
dependence
goes as $P_\tau \cos \beta _a$ where $P_\tau \simeq -0.14$ is the
degree of
tau polarization. Lastly, notice that unlike in K$_{l3}$ decays there
are {\em } no
large
final-state interactions in the ``standard lepton model'', such as an
electromagnetic
rescattering between
the $\rho $ (or its decay products) and the $\nu $.

Two useful generalizations of the above $I\left( E_\rho ,E_{\bar \rho
},\tilde \theta _1,\tilde \theta _2\right) $ for the $\{\rho ^{-},\rho
^{+}\}$
are  \cite{C94a}

\begin{itemize}
\item  to the $\tau \rightarrow a_1\nu $ decay mode in which the
$a_1$has
the opposite $CP$ quantum number to that of the $\rho $ \cite{ai}

\item  to include $\nu _R$ couplings [see next Section].
\end{itemize}

For the $\tau ^{-}\rightarrow a_1^{-}\nu \rightarrow \left( \pi ^{-}\pi
^{-}\pi ^{+}\right) \nu ,\left( \pi ^o\pi ^o\pi ^{-}\right) \nu $ modes,
the
composite-decay-density matrix is given by

$$
\begin{array}{l}
\rho _{hh}=
 \\ \left( 1+h\cos \theta _1^\tau \right) \left[ \sin ^2\omega _1\cos
^2\tilde \theta _1 \\
+ ( 1- \frac 12\sin ^2\omega _1 ) \sin ^2\tilde \theta _1\right]
\\
+
\frac{r_a^2}2\left( 1-h\cos \theta _1^\tau \right) \left[ \left( 1+\cos
^2\omega
_1\right) \cos ^2\tilde \theta _1  \\ +\left( 1+\frac 12\sin ^2\omega
_1\right) \sin
^2\tilde \theta _1 ]  \\ -h\frac{r_a}{\sqrt{2}}\cos \beta _a\sin \theta
_1^\tau \sin 2\omega _1\left[ \cos ^2\tilde \theta _1-\frac 12\sin
^2\tilde
\theta _1\right]
\end{array}
$$
Here $\tilde \theta _1$ specifies the normal to the $\left( \pi ^{-}\pi
^{-}\pi ^{+}\right) $ decay triangle, instead of the $\pi ^{-}$
momentum
direction used for $\tau ^{-}\rightarrow \rho ^{-}\nu $. Also the
Dalitz
plot for $\left( \pi ^{-}\pi ^{-}\pi ^{+}\right) $ has been integrated
over
so that, following Berman \& Jacob, it is not necessary to separate
the
form-factors for $a_1^{-} \rightarrow $ $\left( \pi ^{-}\pi ^{-}\pi
^{+}\right) $%
. The $\left( \pi ^o\pi ^o\pi ^{-}\right) $ mode is similarly treated.

For the two $CP$ tests Table 4 shows that the sensitivity of the
$a_1$mode, versus
the $\rho $ mode,  is about 2 times better for the $r_{a\text{ }}
$ measurement and is about 5 times worse for the $\beta $
measurements.

\subsection{Tests for a (V-A) +  ``something''  \newline Structure}

It is straightforward to include $\nu _R$ and $\bar \nu _L$ couplings
in
spin-correlation functions. The helicity amplitudes for $\tau
^{-}\rightarrow \rho ^{-}\nu _{L,R}$ for both $(V\mp A)$
couplings and $%
m_\nu $ arbitrary are:

For $\nu _L$ so $\lambda _\nu =-\frac 12$,%
$$
\begin{array}{l}
A\left( 0,-\frac 12\right) =g_L\left(
\frac{E_\rho +q_\rho }{m_\rho }\right) \sqrt{m_\tau \left( E_\nu
+q_\rho
\right) } \\ -g_R\left(
\frac{E_\rho -q_\rho }{m_\rho }\right) \sqrt{m_\tau \left( E_\nu -
q_\rho
\right) } \\ A\left( -1,-\frac 12\right) =g_L
\sqrt{2m_\tau \left( E_\nu +q_\rho \right) } \\ -g_R\sqrt{2m_\tau
\left(
E_\nu -q_\rho \right) }.
\end{array}
$$

For $\nu _R$
so $\lambda _\nu =\frac 12$,%
$$
\begin{array}{l}
A\left( 0,\frac 12\right) =-g_L\left(
\frac{E_\rho -q_\rho }{m_\rho }\right) \sqrt{m_\tau \left( E_\nu -
q_\rho
\right) } \\ +g_R\left(
\frac{E_\rho +q_\rho }{m_\rho }\right) \sqrt{m_\tau \left( E_\nu
+q_\rho
\right) } \\ A\left( 1,\frac 12\right) =-g_L
\sqrt{2m_\tau \left( E_\nu -q_\rho \right) } \\ +g_R\sqrt{2m_\tau
\left(
E_\nu +q_\rho \right) }
\end{array}
$$
and $A\left( -1,\frac 12\right) =0$. Note that $g_L,g_R$ denote the
`chirality' of the coupling and $\lambda _\nu =\mp \frac 12$ denote
the
handedness of $\nu _{L,R}$.

The resulting S2SC formulas are also very simple for including $\nu
_R$ and $\bar
\nu
_L$ couplings,

$$
\begin{array}{l}
I\left( E_\rho ,E_{\bar \rho },\tilde \theta _1,\tilde \theta _2\right)
\mid
_{\nu _R,\bar \nu _L}=I_4 \\
+\left( \lambda _R\right) ^2I_4\left( \rho \rightarrow \rho ^R\right)
+\left( \bar \lambda _L\right) ^2I_4\left( \bar \rho \rightarrow \bar
\rho
^L\right)  \\
+\left( \lambda _R\bar \lambda _L\right) ^2I_4\left( \rho \rightarrow
\rho
^R,\bar \rho \rightarrow \bar \rho ^L\right)
\end{array}
$$
where $\lambda _R\equiv $ $\frac{|A\left( 0,\frac 12\right)
|}{|A\left(
0,-\frac 12\right) |},$ $\bar \lambda _L\equiv $ $\frac{|B\left( 0,-\frac
12\right) |}{|B\left( 0,\frac 12\right) |}$ give the moduli's of the $\nu
_R$
and $\bar \nu _L$ amplitudes versus the standard amplitudes. The
corresponding
composite density matrices for $\tau \rightarrow \rho \nu $ with
$\nu _R$
and $\bar \nu _L$ final state particles are given by the substitution
rules:%
$$
\begin{array}{c}
\rho _{hh}^R=\rho _{-h,-h}\left( r_a\rightarrow r_a^R,\beta
_a\rightarrow
\beta _a^R\right)  \\
\bar \rho _{hh}^L=\bar \rho _{-h,-h}\left( r_b\rightarrow r_b^L,\beta
_b\rightarrow \beta _b^L\right)
\end{array}
$$
where the  $\nu _R$ and $\bar \nu _L$ moduli ratios and phase
differences
are defined by $r_a^R\equiv $ $\frac{|A\left( 1,\frac 12\right) |}{%
|A\left( 0,\frac 12\right) |},$ $r_b^L\equiv $ $\frac{|B\left( -1,-\frac
12\right) |}{|B\left( 0,-\frac 12\right) |},\beta _a^R\equiv \phi _1^a-
\phi
_0^{aR},\beta _b^L\equiv \phi _{-1}^b-\phi _0^{bL}$.

The presence of $(V+A)$ couplings in $\tau ^{-}\rightarrow A^{-
}\nu _\tau $
is characterized by the value of the ``chirality parameter'' $\xi
_A\equiv
\frac{|g_L|^2-|g_R|^2}{|g_L|^2+|g_R|^2}=\frac{2Re\left(
v_Aa_A^{*}\right) }{%
|v_A|^2+|a_A|^2}$.  Note that $\xi _A=-\langle h_{\nu _\tau }\rangle
$, twice the
negative of the $\nu _\tau $ helicity, in the special case of only $V$
and $A$
couplings and $m_\nu =0$. Using spin-correlations, the ALEPH
\cite{e2} and the
ARGUS \cite{e1}
collaborations have measured $\xi _A$.  The current world average
is $\xi
_A=1.002\pm 0.032$, see M. Davier's talk \cite{e3}.

\begin{table*}[hbt]
% space before first and after last column: 1.5pc
% space between columns: 3.0pc (twice the above)
\setlength{\tabcolsep}{1.5pc}
% -----------------------------------------------------
% adapted from TeX book, p. 241
\newlength{\digitwidth} \settowidth{\digitwidth}{\rm 0}
\catcode`?=\active \def?{\kern\digitwidth}
% -----------------------------------------------------
\caption{Limits on $\Lambda$ in $GeV$ for Real Coupling
Constants}
\label{tab:real1}
\begin{tabular*}{\textwidth}{@{}l@{\extracolsep{\fill}}rrrr}
\hline
                 & \multicolumn{2}{l}{$\lbrace \rho^{-}, \rho^{+}
\rbrace$ mode}
                 & \multicolumn{2}{l}{$\lbrace a_{1}^{-}, a_{1}^{+}
\rbrace$ mode}
\\
\cline{2-3} \cline{4-5}
                 & \multicolumn{1}{r}{At $M_Z$}
                 & \multicolumn{1}{r}{10, or 4 GeV}
                 & \multicolumn{1}{r}{At $M_Z$}
                 & \multicolumn{1}{r}{10, or 4 GeV}         \\
\hline
{\bf 1st Class Currents}  &                      &                   &
&               \\
$V+A$, for $\xi_A$        & $0.006$      & $0.0012$ & $0.010$ &
$0.0018$ \\
$f_M$, for $\Lambda$   & $214 GeV$ & $1,200$   &  $282$   &
$1,500$ \\
$S$                                   & $306 GeV$ &  $1,700$   &  $64$     &
$345$ \\
$T_5^{+}$                      & $506 GeV$ &  $2,800$   &  $371$   &
$2,000$ \\
{\bf 2nd Class Currents}&                      &                   &
&
\\
$f_E$, for $\Lambda$    & $214 GeV$ & $1,200$    &  $282$   &
$1,500$ \\
$P$                                   & $306 GeV$ &  $1,700$   &  $64$     &
$345$ \\
$T^{+}$                          & $506 GeV$ &  $2,800$   &  $371$   &
$2,000$ \\
\hline
\multicolumn{5}{@{}p{120mm}}{ For $V+A$ only, the entry is
for $\xi_A$.}
\end{tabular*}
\end{table*}

Using both the above 4 variable and the analogous 7 variable S2CS
functions,
we have obtained their associated ideal statistical errors for
measurements of
$\xi_{\rho}$, and $ \xi_{a_1}$ so as to see what are the ``ideal''
limits and to see
whether it is useful to include additional
variables: Using $I_4$, for respectively  $\{\rho ^{-},\rho
^{+}\},\{a_1^{-},a_1^{+}\}$ we find $\sigma =0.006,0.010$ at
$M_Z$, $\sigma
=0.0012,0.0018$ at  $10GeV$, and  $\sigma =0.0013,0.0018$ at
$4GeV.$ This
shows for $\{\rho ^{-},\rho ^{+}\}$ that by using $I_4$, instead of
the
simpler 2 variable $I\left( E_\rho ,E_{\bar \rho }\right) $ spin-
correlation
function, there is about a factor of 8 improvement.  However, using
the 3
additional variables $(\phi, \tilde{\phi_1}, \tilde{\phi_2} )$ in $I_7$
gives less
than a $1\%$ improvement. Even
if in addition the $\tau ^{-}$momentum direction is known via a
SVX detector,
there is only an $\sim 11\%$ improvement. Hence, for $(V+A)$
versus  $(V-A)$
more than 4 variables is not statistically helpful.

\subsection{Tests for Additional Tensorial, \newline Scalar, and
Pseudo-Scalar
Couplings}

Historically in the study of the weak charged-current in muonic and
in
hadronic processes, it has been an important issue to determine the
``complete Lorentz structure'' directly from experiment in a model
independent manner. Here the $I_4$ and $I_7$ functions can be
used for this
purpose to study the $ \tau$ charged-current \cite{a3} since these
functions
depend directly on the 4 helicity amplitudes for $\tau
^{-}\rightarrow \rho ^{-}\nu $ and on the 4 amplitudes for the $CP$-
conjugate
process. We also obtain the associated ``ideal'' sensitivities.

We first consider the ``traditional'' couplings for  $\tau ^{-
}\rightarrow \rho
^{-}\nu $ which characterize the most general Lorentz coupling
$$
\rho _\mu ^{*}\bar u_{\nu _\tau }\left( p\right) \Gamma ^\mu u_\tau
\left(
k\right)
$$
where $k_\tau =q_\rho +p_\nu $. It is convenient to treat the vector
and
axial vector matrix elements separately. We introduce a parameter
$%
\Lambda =$ ``the scale of New Physics''. In effective field theory
this
is the scale at which new particle thresholds are expected to
occur.  In old-fashioned renormalization theory it is the scale at
which the calculational methods and/or the principles of
``renormalization''
breakdown, see for example \cite{th}. While some terms of the
following types do
occur as higher-order perturbative-corrections in the standard model,
such SM
contributions are ``small'' versus the sensitivities of present tests in
$\tau$ physics,
c.f. \cite{d0,d5,t1,d1b,d2,a5}.

In terms of the ``traditional'' tensorial and spin-zero couplings%
$$
\begin{array}{l}
V_{\nu \tau }^\mu \equiv \langle \nu |v^\mu \left( 0\right) |\tau
\rangle
=\bar u_{\nu _\tau }\left( p\right) [g_V\gamma ^\mu  \\
+\frac{f_M}{2\Lambda }\iota \sigma ^{\mu \nu }(k-p)_\nu
+\frac{g_{S^{-}}}{%
2\Lambda }(k-p)^\mu ]u_\tau \left( k\right)
\end{array}
$$
$$
\begin{array}{l}
A_{\nu \tau }^\mu \equiv \langle \nu |a^\mu \left( 0\right) |\tau
\rangle
=\bar u_{\nu _\tau }\left( p\right) [g_A\gamma ^\mu \gamma _5 \\
+\frac{f_E}{2\Lambda }\iota \sigma ^{\mu \nu }(k-p)_\nu \gamma
_5+\frac{%
g_{P^{-}}}{2\Lambda }(k-p)^\mu \gamma _5]u_\tau \left( k\right)
\end{array}
$$
Notice that $\frac{f_M}{2\Lambda }=$ a ``tau weak magnetism''
type
coupling, and $\frac{f_E}{2\Lambda }=$ a ``tau weak electricity''
type
coupling. Both the scalar $g_{S^{-}}$ and pseudo-scalar  $g_{P^{-
}}$%
couplings do not contribute for  $\tau ^{-}\rightarrow \rho ^{-}\nu $
since $%
\rho _\mu ^{*}q^\mu =0$, nor for  $\tau ^{-}\rightarrow a_1^{-}\nu
$.

\begin{table*}[hbt]
% space before first and after last column: 1.5pc
% space between columns: 3.0pc (twice the above)
\setlength{\tabcolsep}{1.5pc}
% -----------------------------------------------------
% adapted from TeX book, p. 241
\newlength{\digitwidth} \settowidth{\digitwidth}{\rm 0}
\catcode`?=\active \def?{\kern\digitwidth}
% -----------------------------------------------------
\caption{``Reality structure'' of $J^{\mu}_{Lepton}$ current's form
factors:}
\label{tabr}
\begin{tabular*}{\textwidth}{@{}l@{\extracolsep{\fill}}|rr|r}
\hline
 {\bf Form Factor:}             & Class I Current & Class II Current&
{\bf T
invariance} \\
\hline
$V,A,f_{M},P^{-}$           & Real parts          & Imaginary parts &
$Re\neq 0,
Im=0$  \\
$f_{E},S^{-}$                    & Imaginary parts & Real parts          &
$Re\neq 0,
Im=0$  \\
\hline
\end{tabular*}
\end{table*}

\begin{table*}[hbt]
% space before first and after last column: 1.5pc
% space between columns: 3.0pc (twice the above)
\setlength{\tabcolsep}{1.5pc}
% -----------------------------------------------------
% adapted from TeX book, p. 241
\newlength{\digitwidth} \settowidth{\digitwidth}{\rm 0}
\catcode`?=\active \def?{\kern\digitwidth}
% -----------------------------------------------------
\caption{Limits on $\Lambda$ in $GeV$ for Pure Imaginary
Coupling Constants}
\label{tab:imag1}
\begin{tabular*}{\textwidth}{@{}l@{\extracolsep{\fill}}rrrr}
\hline
                 & \multicolumn{2}{l}{$\lbrace \rho^{-}, \rho^{+}
\rbrace$ mode}
                 & \multicolumn{2}{l}{$\lbrace a_{1}^{-}, a_{1}^{+}
\rbrace$ mode}
\\
\cline{2-3} \cline{4-5}
                 & \multicolumn{1}{r}{At $M_Z$}
                 & \multicolumn{1}{r}{10, or 4 GeV}
                 & \multicolumn{1}{r}{At $M_Z$}
                 & \multicolumn{1}{r}{10, or 4 GeV}         \\
\hline
{\bf 1st Class Currents:}     &                          &                  &
&
\\
$V+A$, for $\xi_A$             & $0.006$          & $0.0012$ & $0.010$
& $0.0018$
\\
$f_M$, for $(\Lambda)^2$ & $(12GeV)^2$ & $(28)^2$ &
$(15)^2$& $(34)^2$ \\
$S$                                         & $(14GeV)^2$ & $(33)^2$ & $(
6)^2$ & $(13)^2$
\\
$T_5^{+}$                            & $(22GeV)^2$ & $(50)^2$ &
$(18)^2$& $(42)^2$
\\
{\bf 2nd Class Currents:}     &                          &                  &
&
\\
$f_E$, for $(\Lambda)^2$   & $(12GeV)^2$ & $(28)^2$ &
$(15)^2$& $(34)^2$ \\
$P$                                         & $(14GeV)^2$ & $(33)^2$ & $(
6)^2$ & $(13)^2$
\\
$T^{+}$                                 & $(22GeV)^2$ & $(50)^2$ &
$(18)^2$& $(42)^2$
\\
\hline
\multicolumn{5}{@{}p{120mm}}{ For $V+A$ only, the entry is
for $\xi_A$.}
\end{tabular*}
\end{table*}

By Lorentz invariance, there is the equivalence theorem that for the
vector
current%
$$
\begin{array}{cc}
S\approx V+f_M, & T^{+}\approx V+S^{-}
\end{array}
$$
and for the axial-vector current
$$
\begin{array}{cc}
P\approx A+f_E, & T_5^{+}\approx A+P^{-}
\end{array}
$$
where
$$
\begin{array}{lcc}
\Gamma _V^\mu =g_V\gamma ^\mu +
\frac{f_M}{2\Lambda }\iota \sigma ^{\mu \nu }(k-p)_\nu  \\ +
\frac{g_{S^{-}}}{2\Lambda }(k-p)^\mu +\frac{g_S}{2\Lambda
}(k+p)^\mu  \\
+%
\frac{g_{T^{+}}}{2\Lambda }\iota \sigma ^{\mu \nu }(k+p)_\nu
\end{array}
$$

$$
\begin{array}{lcc}
\Gamma _A^\mu =g_A\gamma ^\mu \gamma _5+
\frac{f_E}{2\Lambda }\iota \sigma ^{\mu \nu }(k-p)_\nu \gamma _5
\\ +
\frac{g_{P^{-}}}{2\Lambda }(k-p)^\mu \gamma
_5+\frac{g_P}{2\Lambda }%
(k+p)^\mu \gamma _5 \\ +\frac{g_{T_5^{+}}}{2\Lambda }\iota
\sigma ^{\mu \nu
}(k+p)_\nu \gamma _5
\end{array}
$$

The matrix elements of the divergences of these charged-currents
are%
$$
\begin{array}{lcc}
(k-p)_\mu V^\mu =[g_V(m_\nu -m_\tau ) \\
+
\frac{g_{S^{-}}}{2\Lambda }q^2+\frac{g_S}{2\Lambda }(m_\nu
^2-m_\tau ^2)
\\ +%
\frac{g_{T^{+}}}{2\Lambda }(q^2-[m_\tau -m_\nu ]^2)]\bar u_\nu
u_\tau
\end{array}
$$
$$
\begin{array}{lcc}
(k-p)_\mu A^\mu =[g_A(m_\nu +m_\tau ) \\
+
\frac{g_{P^{-}}}{2\Lambda }q^2+\frac{g_P}{2\Lambda }(m_\nu
^2-m_\tau ^2)
\\ +%
\frac{g_{T_5^{+}}}{2\Lambda }(q^2-m_\tau ^2+m_\nu ^2)]\bar
u_\nu \gamma
_5u_\tau
\end{array}
$$
Both the weak magnetism  $\frac{f_M}{2\Lambda }$and the weak
electricty $%
\frac{f_E}{2\Lambda }$ terms are divergenceless. On the other
hand, since $%
q^2=m_\rho ^2$,  when $m_\nu =m_\tau $ there are non-vanishing
terms due to
the couplings $S^{-},T^{+},A,P^{-},T_5^{+}$.

Table 5 gives the limits on these additional couplings assuming a
$(V-A)+$%
``something'' structure for the tau charged-current. Real coupling
constants
are assumed. Notice that at $M_Z$ the scale of $\Lambda \approx
$few $\
100GeV
$ can be probed; and at $10GeV$ or at $4GeV$ the scale of $1-
2TeV$ can be
probed.

In compiling the entries in Table 5, we have used the idea of 1st and
2nd
class currents \cite{sc1}. This is suggested by a 3rd-family
perspective of a possible ``$\tau \leftrightarrow \nu _\tau $
symmetry''
in the structure of the tau lepton currents. At the level of the masses,
this is a badly broken symmetry but it might still be relevant to 3rd-
family
currents.  Recall that $\frac{m_b}{m_t}\sim \frac 5{174}\sim 3\%$,
and $%
\frac{m_\nu }{m_\tau }<\frac{24}{1777}\sim 1.7\%$ so this
symmetry is badly
broken for the 3rd family. However, for the other leptons this
symmetry is
empirically (and also for Harari's theoretical values) more strongly
broken since
$\frac{m_{\nu
_e}}{m_e}<10^{-5}$, and $\frac{m_{\nu _\mu }}{m_\mu
}<0.15\%$.

\begin{table*}[hbt]
% space before first and after last column: 1.5pc
% space between columns: 3.0pc (twice the above)
\setlength{\tabcolsep}{1.5pc}
% -----------------------------------------------------
% adapted from TeX book, p. 241
\newlength{\digitwidth} \settowidth{\digitwidth}{\rm 0}
\catcode`?=\active \def?{\kern\digitwidth}
% -----------------------------------------------------
\caption{Contributions to $A(\lambda_{\rho},\lambda_{\nu})$ from
the general
$\rho^{*}_{\alpha} ( \overline{u}_{\nu} \Gamma^{\alpha}
u_{\tau} )$ coupling
versus $m_{\nu} \rightarrow 0$ limit:}
\label{tabc}
\begin{tabular*}{\textwidth}{@{}l@{\extracolsep{\fill}}|rrrr}
\hline
                 & \multicolumn{1}{r}{$V-A$}
                 & \multicolumn{1}{r}{$V+A$}
                 & \multicolumn{1}{r}{$S+P$}
                 & \multicolumn{1}{r}{$S-P$}         \\
\hline
{\bf $\nu_L$ Helicity Amplitudes:}   &                    &
&
&         \\
$A(0,-1/2)$                                           &  $ \bullet $ &  $ \circ
$   &
$ \bullet $  &
$ \circ $ \\
$A(-1,-1/2)$                                         &  $ \bullet $ &  $  \circ
$  &
&         \\
{\bf $\nu_R$ Helicity Amplitudes:}  &                    &
&
&        \\
$A(0,1/2)$                                            &  $ \circ  $  &  $
\bullet $
& $ \circ  $   &
$ \bullet $ \\
$A(1,1/2)$                                            &  $  \circ $  &  $
\bullet $
&
&         \\
\hline
\multicolumn{5}{@{}p{120mm}}{Here $ \bullet = $ `` occurs even
if $m_{\nu}
= 0$'' and $ \circ = $ `` requires $m_{\nu } \neq 0$.'' \newline The
$T^{+} \pm
T^{+}_5$ couplings are respectively mixtures of the $V \mp A$ and
the $S\pm
P$. So, the $T^{+} \pm T^{+}_5$ have the same columnar entries
as that shown
for the $V \mp A$.}
\end{tabular*}
\end{table*}

\begin{table*}[hbt]
% space before first and after last column: 1.5pc
% space between columns: 3.0pc (twice the above)
\setlength{\tabcolsep}{1.5pc}
% -----------------------------------------------------
% adapted from TeX book, p. 241
\newlength{\digitwidth} \settowidth{\digitwidth}{\rm 0}
\catcode`?=\active \def?{\kern\digitwidth}
% -----------------------------------------------------
\caption{``Chiral Couplings'':  Limits on $\Lambda$ in $GeV$ for
Real Coupling
Constants}
\label{tab:creal1}
\begin{tabular*}{\textwidth}{@{}l@{\extracolsep{\fill}}rrrr}
\hline
                 & \multicolumn{2}{l}{$\lbrace \rho^{-}, \rho^{+}
\rbrace$ mode}
                 & \multicolumn{2}{l}{$\lbrace a_{1}^{-}, a_{1}^{+}
\rbrace$ mode}
\\
\cline{2-3} \cline{4-5}
                 & \multicolumn{1}{r}{At $M_Z$}
                 & \multicolumn{1}{r}{10, or 4 GeV}
                 & \multicolumn{1}{r}{At $M_Z$}
                 & \multicolumn{1}{r}{10, or 4 GeV}         \\
\hline
$V+A$, for $\xi_A$                    & $0.006$         & $0.0012$ &
$0.010$ &
$0.0018$ \\
$S+P$, for $\Lambda$                & $310 GeV$   & $1,700$   &  $
64$     & $
350$ \\
$S-P$, for $(\Lambda)^2$          &$(11GeV)^2$& $(25)^2$ &
$(4)^2$  &
$(7)^2,(10)^2$ \\
$f_M+f_E$, for $\Lambda$      & $210 GeV$   & $1,200$   &
$280$   & $1,500$
\\
$f_M-f_E$, for $(\Lambda)^2$&$(9GeV)^2$  & $(20)^2$ &
$(10)^2$& $(24)^2$
\\
\hline
\multicolumn{5}{@{}p{120mm}}{ For the $\rho$ and $a_1$
modes, the
$T^{+}+T_5^{+}$ coupling is equivalent to the $V-A$ coupling;
and $T^{+}-
T_5^{+}$ is equivalent to $V+A$.}
\end{tabular*}
\end{table*}

\begin{table*}[hbt]
% space before first and after last column: 1.5pc
% space between columns: 3.0pc (twice the above)
\setlength{\tabcolsep}{1.5pc}
% -----------------------------------------------------
% adapted from TeX book, p. 241
\newlength{\digitwidth} \settowidth{\digitwidth}{\rm 0}
\catcode`?=\active \def?{\kern\digitwidth}
% -----------------------------------------------------
\caption{``Chiral Couplings'':  Limits on $\Lambda$ in $GeV$ for
Pure Imaginary
Coupling Constants}
\label{tab:cimag1}
\begin{tabular*}{\textwidth}{@{}l@{\extracolsep{\fill}}rrrr}
\hline
                 & \multicolumn{2}{l}{$\lbrace \rho^{-}, \rho^{+}
\rbrace$ mode}
                 & \multicolumn{2}{l}{$\lbrace a_{1}^{-}, a_{1}^{+}
\rbrace$ mode}
\\
\cline{2-3} \cline{4-5}
                 & \multicolumn{1}{r}{At $M_Z$}
                 & \multicolumn{1}{r}{10, or 4 GeV}
                 & \multicolumn{1}{r}{At $M_Z$}
                 & \multicolumn{1}{r}{10, or 4 GeV}         \\
\hline
$V+A$, for $\xi_A$                    & $0.006$         & $0.0012$ &
$0.010$ &
$0.0018$ \\
$S+P$, for$(\Lambda)^2$          &$(11GeV)^2$& $(25)^2$ &
$(4)^2$  &
$(10)^2$ \\
$S-P$, for $(\Lambda)^2$          &$(11GeV)^2$& $(25)^2$ &
$(4)^2$  &
$(7)^2,(10)^2$ \\
$f_M+f_E$, for$(\Lambda)^2$&$(9GeV)^2$  & $(20)^2$ &
$(10)^2$& $(24)^2$
\\
$f_M-f_E$, for $(\Lambda)^2$&$(9GeV)^2$  & $(20)^2$ &
$(10)^2$& $(24)^2$
\\
\hline
\multicolumn{5}{@{}p{120mm}}{For the $\rho$ and $a_1$
modes, the
$T^{+}+T_5^{+}$ coupling is equivalent to the $V-A$ coupling;
and $T^{+}-
T_5^{+}$  is equivalent to the $V+A$.}
\end{tabular*}
\end{table*}

We assume that the effective charged-current
${J_{Lepton}}^{Charged}$ is
Hermitian and has such an
SU(2) symmetry, so that we can identify the $\nu _\tau $ and the
$\tau ^{-}$%
spinors. Thereby, we obtain for the ``traditional couplings'' and real
form factors
that the ``Class I'' couplings are $V,A,f_M,P^{-}$, and that the
``Class
II'' couplings are  $f_E,S^{-}$ if we define  $J_{Lepton}^\mu =$
$J_I^\mu +$ $%
J_{II}^\mu $ where for $U=\exp (\iota \pi I_2)$%
$$
\begin{array}{cc}
(J_I^\mu )^{\dagger }=-UJ_I^\mu U^{-1} & First \\
(J_{II}^\mu )^{\dagger }=UJ_{II}^\mu U^{-1} & Second
\end{array}
Class
$$

This classification is particularly useful in considering the reality
structure of the charged-current \cite{sc2}. As show in Table 6 there
is a ``clash''
between the ``Class I and Class II'' structures and
the consequences of
time-reversal invariance. In particular, there are the useful theorems
that (a) ($\tau \leftrightarrow \nu _\tau $ symmetry) + ($T$
invariance) $%
\Longrightarrow $ Class II currents are absent, (b) ($\tau
\leftrightarrow
\nu _\tau $ symmetry) + (existence of $J_I^\mu $ and $J_{II}^\mu
$) $%
\Longrightarrow $ violation of $T$ invariance, and (c) (existence of
$%
J_{II}^\mu $) +  ($T$ invariance) $\Longrightarrow $($\tau
\leftrightarrow \nu _\tau $ symmetry) in $J_{Lepton}^\mu $ is
broken.

Table 7 shows the limits on such couplings assuming a pure-
imaginary
coupling constant. In the case of $(V-A)$ the limits on the $ \beta $'s
in Sec. 2.2
cover this situation. Notice that the limits here are in $ ( \Lambda
)^2$and are $
\Lambda \sim$ few $10GeV$'s because, unlike for  Table 6, this is
not due to an
interference effect in the S2SC functions.

Besides the 3rd-family perspective of a possible $\tau \leftrightarrow
\nu _\tau $ symmetry, it is also instructive to consider the situation
from the
perspective of ``Chiral Combinations'' of the various
Lorentz couplings, see Table 8.  Note that the $S\pm P$ couplings
do not
contribute
to the transverse $\rho $ or $a_{1\text{ }}$transitions. Tables 9 and
10 give the
limits on $\Lambda $ in the case of purely real and
imaginary coupling constants for these ``Chiral Couplings''.

The results in the tables in this section easily follow from the
dependence
of the helicity amplitudes for $\tau ^{-}\rightarrow \rho ^{-}\nu $ on
the
presence of $(S\pm P)$ couplings with $m_\nu $ arbitrary:

$$
\begin{array}{l}
A(0,-\frac 12)=g_{S+P}(
\frac{m_\tau }{2\Lambda })\frac{2q_\rho }{m_\rho }\sqrt{m_\tau
(E_\rho
+q_\rho )} \\ +g_{S-P}(\frac{m_\tau }{2\Lambda })\frac{2q_\rho
}{m_\rho }%
\sqrt{m_\tau (E_\rho -q_\rho )}
\end{array}
$$
$$
A(-1,-\frac 12)=0
$$

$$
\begin{array}{l}
A(0,\frac 12)=g_{S+P}(
\frac{m_\tau }{2\Lambda })\frac{2q_\rho }{m_\rho }\sqrt{m_\tau
(E_\rho
-q_\rho )} \\ +g_{S-P}(\frac{m_\tau }{2\Lambda })\frac{2q_\rho
}{m_\rho }%
\sqrt{m_\tau (E_\rho +q_\rho )}
\end{array}
$$
$$
A(1,\frac 12)=0
$$

\begin{table*}[hbt]
% space before first and after last column: 1.5pc
% space between columns: 3.0pc (twice the above)
\setlength{\tabcolsep}{1.5pc}
% -----------------------------------------------------
% adapted from TeX book, p. 241
\newlength{\digitwidth} \settowidth{\digitwidth}{\rm 0}
\catcode`?=\active \def?{\kern\digitwidth}
% -----------------------------------------------------
\caption{ Limits on $\nu_R$ and $\={\nu}_L$ couplings in terms of
the ratios of
the moduli of the helicity amplitudes:}
\label{tab:nu1}
\begin{tabular*}{\textwidth}{@{}l@{\extracolsep{\fill}}rrrr}
\hline
                 & \multicolumn{2}{l}{$\lbrace \rho^{-}, \rho^{+}
\rbrace$ mode}
                 & \multicolumn{2}{l}{$\lbrace a_{1}^{-}, a_{1}^{+}
\rbrace$ mode}
\\
\cline{2-3} \cline{4-5}
                 & \multicolumn{1}{r}{At $M_Z$}
                 & \multicolumn{1}{r}{10, or 4 GeV}
                 & \multicolumn{1}{r}{At $M_Z$}
                 & \multicolumn{1}{r}{10, or 4 GeV}         \\
\hline
$\lambda_R$
&$(7\%)^2$&$(3\%)^2$&$(13\%)^2$&$(6\%)^2$ \\
$\lambda_R r_a^R$
&$(7\%)^2$&$(3\%)^2$&$(10\%)^2$&$(4\%)^2$ \\
\hline
\multicolumn{5}{@{}p{120mm}}{Elements of ``error matrix'' are
given in the
next table.}
\end{tabular*}
\end{table*}

\begin{table*}[hbt]
% space before first and after last column: 1.5pc
% space between columns: 3.0pc (twice the above)
\setlength{\tabcolsep}{1.5pc}
% -----------------------------------------------------
% adapted from TeX book, p. 241
\newlength{\digitwidth} \settowidth{\digitwidth}{\rm 0}
\catcode`?=\active \def?{\kern\digitwidth}
% -----------------------------------------------------
\caption{Elements of error matrix for limits on $\nu_R$ and
$\={\nu}_L$
couplings in terms of  the helicity amplitudes for respectively $\tau
\rightarrow
\rho \nu$, and $\tau \rightarrow a_1 \nu$:}
\label{tab:nu2}
\begin{tabular*}{\textwidth}{@{}l@{\extracolsep{\fill}}rrrr}
\hline
                 & \multicolumn{2}{l}{$\lbrace \rho^{-}, \rho^{+}
\rbrace$ mode}
                 & \multicolumn{2}{l}{$\lbrace a_{1}^{-}, a_{1}^{+}
\rbrace$ mode}
\\
\cline{2-3} \cline{4-5}
                 & \multicolumn{1}{r}{At $M_Z$}
                 & \multicolumn{1}{r}{10, or 4 GeV}
                 & \multicolumn{1}{r}{At $M_Z$}
                 & \multicolumn{1}{r}{10, or 4 GeV}         \\
\hline
{\bf Diagonal elements:}                       &                     &
&
&            \\
$ a=\lambda_R$                                      &$(8\%)^2$  &$(4\%)^2$
&$(18\%)^2$&$(8\%)^2,(9\%)^2$ \\
$ b=\lambda_R r_a^R$                          &$(8\%)^2$   &$(4\%)^2$
&$(18\%)^2$&$(8\%)^2,(9\%)^2$ \\
$ c = $                                                      &
&
&
&         \\
$(\lambda_R)^2r_a^R
\cos{\beta^R}$&$(13\%)^2$&$(6\%)^2,(10\%)^2$&$(41\%)^2$&$(
20\%)^2,(24\
%)^2$ \\
{\bf Correlations:}                                 &                     &
&
&          \\
$ \rho_{ab}$                                           &$-0.75$       &$-0.77$
&$-0.95$
&$-0.96, -0.97$ \\
$ \rho_{ac}$                                           &$-0.27$       &$-
0.17,0.06$&$-0.56$
&$0.029,0.019$ \\
$ \rho_{bc}$                                           &$0.085$
&$.017,0.003$&$0.04$
&$-0.041,-0.026$    \\
\hline
\multicolumn{5}{@{}p{120mm}}{}
\end{tabular*}
\end{table*}

Finally, as shown in Tables 11 and 12,   the helicity amplitudes
themselves
provide a simple framework for
characterizing a ``complete measurement'' of the $\tau ^{-
}\rightarrow \rho
^{-}\nu $:  In the case that only $\nu _L$ coupling's exist (compare
the discussion
in
Sec.2.2), there are only 2 amplitudes.  So 3 measurements (of
$r_a,\beta _a,$and
$%
|A(0,-\frac 12)|$ via $\{\rho ^{-},B^{+}\}\mid _{B\neq \rho }$) will
provide
a ``complete measurement''. In the case that $\nu _R$ coupling's also
exist, then
there are 2 more amplitudes, $A(0,\frac 12)$ and $A(1,\frac 12)$.
Then there are
2 additional CP tests: $r_a^R=r_b^L$ and $\beta _a^R=\beta _b^L$.
Then to
achieve
an ``almost'' complete measurement, 3 additional quantities must be
determined, e.g. by the $I_4$ S2SC function:  $r_a^R,\beta _a^R$
and $%
\lambda _R\equiv \frac{|A(0,\frac 12)|}{|A(0,-\frac 12)|}$. However,
to also
measure
the relative phase of the $\nu _L$ and $\nu _R$ amplitudes, $\beta
_a^o\equiv
\phi _o^{aR}-\phi _o^a$ or $\beta _a^1\equiv \phi _1^a-\phi _{-
1}^a$, requires,
e.g., the occurrence of a common final state which arises from both
$\nu _L$ and
$\nu _R$.

\section{``ADDITIONAL STRUCTURE'' IN \newline
NEUTRAL-CURRENT COUPLINGS?}

Here, our discussion is brief and we refer the reader to other talks at
this
conference \cite{t1,t2} on the Lorentz structure of neutral-current
couplings.
Again it is important to perform a complete systematic search for
possible
additional $ S,P,$ and $T$ couplings, with either real or imaginary
coupling
constants, and to determine the experimental bounds on the
associated scales
$\Lambda$ which probe for possible new physics.  The helicity
formalism in the
context of ``beam-referenced spin-correlation'' functions provides a
simple
framework for such an investigation since simple BRSC functions
exist \cite{a5}
and the relevant helicity amplitudes are simply expressed in terms of
the most
general Lorentz couplings (see Sec. 3 of Ref. \cite{v1} ).

To obtain BRSC functions,  the simpler $I(E_1,E_2)$
spin-correlation function is generalized by including the polar and
azimuthal angles of the incident $e^{-}$ beam versus the final $\pi
^{-}$and $%
\pi ^{+}$ momenta [e.g. \^z along the final $\pi ^{-}$ with the
orthogonal
\^x in (\^z,$\pi ^{+}$) half-plane. Then $\theta _{beam,}\phi
_{beam}$ describes
the
incident $e^{-}$  in this \^x , \^y, \^z right-handed coordinate
system].

By now, there is a considerable literature on tests for such
anomalous couplings, in
particular for electromagnetic and/or weak dipole moments in
$\gamma
^{*}\rightarrow \tau ^{-}\tau ^{+}$ \cite{d0,d5} and in
$Z^{o}\rightarrow \tau
^{-}\tau ^{+}$ \cite{d1,d1b,a5,d2,d3,d4,p1,d6}.   There are
experimental bounds
on the real and imaginary parts of a weak dipole moment
$\tilde{d}(q^2) $ from
the ALEPH and OPAL collaborations \cite{d4e,t2}.

For $Z^o,$ or $\gamma ^{*}\rightarrow \tau ^{-}\tau ^{+}$, there
are 4
independent (complex) helicity amplitudes $T(\pm \mp )$ and
$T(\pm \pm )$
where the matrix element%
$$
\langle \Theta _\tau ,\Phi _\tau |\text{J=1},\text{M}\rangle
=\text{D}%
_{M\lambda }^{1*}(\Phi _\tau ,\Theta _\tau ,-\Phi _\tau )\text{
T}\left(
\lambda _1,\lambda _2\right)
$$
Table 13 lists the discrete symmetry relations for these production
amplitudes. In principle, these amplitudes can be completely
determined by
measurement of the BRSC function for the process $%
e^{-}e^{+}\rightarrow Z^o,$ or $\gamma ^{*}\rightarrow \tau ^{-
}\tau
^{+}\rightarrow (\pi ^{-}\nu )(\pi ^{+}\bar \nu )$.

\begin{table}[hbt]
% space before first and after last column: 1.5pc
% space between columns: 3.0pc (twice the above)
\setlength{\tabcolsep}{.25pc}
% -----------------------------------------------------
% adapted from TeX book, p. 241
\newlength{\digitwidth} \settowidth{\digitwidth}{\rm 0}
\catcode`?=\active \def?{\kern\digitwidth}
% -----------------------------------------------------
\caption{Discrete symmetry relations for production \newline
amplitudes for
$Z^{o}, \gamma^{*} \rightarrow \tau^{-} \tau^{+} $:}
\label{tabp}
\begin{tabular}{lrr}
\hline

\multicolumn{1}{c}{\bf
If Decay }
                                                                  &
\multicolumn{1}{r}{\bf Then}
                                                                  &
\multicolumn{1}{r}{}
\\
\hline
{CP invar.}       &                                 &   $T(++) = T(--) $
\\
{P invar.}         &   $T(+-) = T(-+)$,  &   $T(++) = T(--) $
\\
{C invar.}         &   $T(+-) = T(-+)$   &
\\
\hline
\multicolumn{3}{@{}p{60mm}}{ P and  C invariances can be
tested for in
$\gamma^{*} \rightarrow \tau^{-} \tau^{+} $.}
\end{tabular}
\end{table}

Table 14 lists the 4 distinct tests for $CP/\tilde T_{FS}$
violation in these production amplitudes. Analogous to the Michel
parameters
in $\mu $ decay, here there are 9 vertex intensity parameters which
characterize possible $CP/\tilde T_{FS}$ violation; so there are 5
vertex
intensity relations which can be used as consistency checks if such a
violation is observed \cite{a5}.

\begin{table}[hbt]
% space before first and after last column: 1.5pc
% space between columns: 3.0pc (twice the above)
\setlength{\tabcolsep}{0pc}
% -----------------------------------------------------
% adapted from TeX book, p. 241
\newlength{\digitwidth} \settowidth{\digitwidth}{\rm 0}
\catcode`?=\active \def?{\kern\digitwidth}
% -----------------------------------------------------
\caption{4 distinct tests for  $CP$ / $\~{T_{FS}}$ violation in the
production
process $Z^{o}, \gamma^{*} \rightarrow \tau^{-} \tau^{+}$:}
\label{tabb}
\begin{tabular}{lr}
\hline

\multicolumn{1}{c}{\bf
If symmetry }
                                                                  &
\multicolumn{1}{c}{\bf Then} \\
\hline
                              &                                   \\
{$\~{T_{FS}}$} &                                    \\
                              &  $\beta_{+-} \equiv \phi_{+-} - \phi_{-+}$ \\
                              &    $ = 0$  \hspace{3.4pc} \\
                              &  $\beta_{++} \equiv \phi_{++} - \phi_{--}$\\
                              &    $ = 0$   \hspace{3.4pc} \\
                              &  $2 \beta_{o} \equiv \phi_{++} + \phi_{--} -
\phi_{+-} -
\phi_{-+}$ \\
                              &    $ = 0$  \hspace{8.4pc} \\
{$CP$}                &                                                  \\
                              &  $\beta_{++}  = 0   $     \hspace{3.4pc}
\\
                              &  $\lambda = 0 $  \hspace{3.4pc}   \\
\hline
\multicolumn{2}{@{}p{75mm}}{$\lambda \equiv (|T(++)| - |T(--)|)
/ (|T(++)| +
|T(--)|)$}
\end{tabular}
\end{table}

Table 15 lists the ideal sensitivity for a complete
measurement of these production amplitudes.

\begin{table}[hbt]
% space before first and after last column: .3pc
% space between columns: 3.0pc (twice the above)
\setlength{\tabcolsep}{.3pc}
% -----------------------------------------------------
% adapted from TeX book, p. 241
\newlength{\digitwidth} \settowidth{\digitwidth}{\rm 0}
\catcode`?=\active \def?{\kern\digitwidth}
% -----------------------------------------------------
\caption{Errors for a complete measurement of the $Z^{o}$, or $
\gamma^{*}
\rightarrow \tau^{-} \tau^{+} $ helicity amplitudes:}
\label{tabb1}
\begin{tabular}{lr}
\hline
$(\tau - \gamma^{*} - \tau )$ vertex at $10 GeV$                    &   at
$4 GeV$ \\
\hline
$|t(-+)| = |t(+-)| $ to $0.4\%$
&
$0.4\%$     \\
$|t(++)| = |t(--)| $ to $0.8\%$
&
$0.2\%$     \\
$\beta_{+-}   $  to $ 0.5^{o}$
&
$0.7^{o}$ \\
$\beta_{o}   $  to $ 1^{o}$
&
$0.7^{o}$ \\
$\beta_{++}   $  to $ 4^{o}$
&
$1.5^{o}$  \\
$\sigma (\lambda ) = 0.007$
&
$0.024$      \\
\hline
$( \tau - Z^{o} - \tau )$ vertex at $M_Z$                                  &
\\
\hline
$|T(-+)|  $ to $1\%$
&
\\
$|T(+-)|  $ to $1\%$
&
\\
$\beta_{+-}   $  to $ 3^{o}$
&
\\
\hline
\multicolumn{2}{@{}p{75mm}}{Without experimental surprises,
the $Z^{o}$
helicity-changing
$T(++), T(--)$ will be unmeasured since $ \sigma ( |T(--)| ) /  |T(--)|
=6$.}
\end{tabular}
\end{table}

As noted below Table 15, unless experimental surprises are
discovered, at LEP the
``helicity-changing neutral-current'' amplitudes $T(\pm \pm )$ will
remain
unmeasured. But experimentally ``Is there something where the
Standard Model
says there is
nothing?'' Here again for the $\tau $ lepton the situation is less
unfavorable than
for the other charged leptons since
$$
\frac{|T(--)|}{|T(-+)|}\simeq
\sqrt{2}\frac{v_fm_{lepton}}{a_fM_Z}\simeq
10^{-3}
$$
for $\tau $, but is $10^{-4}$ for $\mu $, $10^{-6}$ for $e$.

\begin{table}[hbt]
% space before first and after last column: 1.5pc
% space between columns: 3.0pc (twice the above)
\setlength{\tabcolsep}{.3pc}
% -----------------------------------------------------
% adapted from TeX book, p. 241
\newlength{\digitwidth} \settowidth{\digitwidth}{\rm 0}
\catcode`?=\active \def?{\kern\digitwidth}
% -----------------------------------------------------
\caption{Ideal statistical errors with respect to C and P tests for the
production
process $ \gamma^{*} \rightarrow \tau^{-} \tau^{+} $:}
\label{tabc}
\begin{tabular}{lrr}
\hline
 { $( \tau - \gamma^{*} - \tau )$ vertex}
&{ $10 GeV$}
&  { $4 GeV$} \\
\hline
``C or P good'' \hspace{1mm} \Longrightarrow  $\alpha_H = 0$ &
$0.002$       &
$0.002$         \\
\hspace{25mm} \Longrightarrow  $\omega = \eta  $                     &
$3\%$
&   $2\%$    \\
``P or CP good'' \Longrightarrow  $\zeta = 1 $                               &
$6\%$
&   $1\%$           \\
\hline
\multicolumn{3}{@{}p{75mm}}{Note that $\alpha_H = P_{\tau}$,
the tau
polarization.}
\end{tabular}
\end{table}

Also at $10GeV$, and at $4GeV$, there are 3 tests for $C$ and $P$
invariances as shown with their associated sensitivities in Table 16.
The vertex
intensity parameters are defined in Ref. \cite{a5}:
$$
\omega ,\eta \equiv (Re[\{T(++)+T(--)\}T^{*}(\mp \pm )])/N
$$
$$
\zeta \equiv 2Re[T(++)T^{*}(--)]/(|T(++)|^2+|T(--)|^2)
$$
are mainly functions of the helicity-changing amplitudes.  But,
$$
\alpha _H\equiv P_{\tau}\equiv  [|T(+-)|^2-|T(-+)|^2]/N
$$
depends mainly on the helicity-conserving
amplitudes. Here $N\equiv \sum |T(\lambda _{1,}\lambda _2)|^2$.

\section{CONCLUSIONS}

(1)\quad In searching for new 3rd-family phenomena in the context
of large
and growing  ($\tau ^{-}\tau ^{+})$ data samples, we have two
powerful tools

\begin{itemize}
\item  tau polarimetry

\item  $(\tau ^{-}\tau ^{+})$ spin correlations.
\end{itemize}
In addition, in the near future there is the very exciting possiblity
that
longitudinally polarized beams will be available in a ``Tau Charm
Factory''
\cite{ch1,ch2}.

\negthinspace \ While the analogous tools \cite{a4} can also
someday be used in
$%
(t\,\bar t)$ physics, there is at present a crucial  difference. Until, at
least, near the
end of this century, there will only be large data samples
for the other than ``$t$'' members of the 3rd-family.

(2)\quad There are observables for tests for ``Additional Structure'' in

\begin{itemize}
\item  tau charged-current couplings
\end{itemize}

\begin{itemize}
\item  tau neutral-current couplings.
\end{itemize}

For instance, in charged-current couplings
\newpage
\newline

$\qquad f_M(q^2)$ tau weak magnetism

$\qquad f_E(q^2)$ tau weak electricity

can be probed to new physics scales of

\quad$\Lambda _{RealCoupling}$\quad   $1.2-1.5TeV$ at $10,$ or
$4GeV$

\quad$\Lambda _{Imag.Coupling}$\quad $28-34GeV$ at $10,$ or
$4GeV$.

For neutral-current couplings, there are
\newline
(i) tests for ``anomalous couplings'' in $Z^o\rightarrow \tau
^{-}\tau ^{+}$
\newline
(ii)$C$ and $P$ tests in $\gamma ^{*}\rightarrow \tau ^{-}\tau
^{+}$

In principle, by spin-correlation techniques the Lorentz stucture of
the neutral-
current couplings can be
completely determined and that of the charge-current couplings can
almost be
completely determined from, for instance,  the $\{\rho ^{-},\rho
^{+}\}$and $%
\{a_1^{-},a_1^{+}\}$ modes.

(3)\quad There are observables for testing for $T$ and/or $CP$
violation in tau
decays, and in the $Z^o$ and $\gamma ^{*}\rightarrow \tau ^{-}\tau
^{+}$
production processes.

\section*{ACKNOWLEDGEMENTS}
For helpful discussions, we thank participants at this conference as
well as
theorists and experimentalists at  Cornell, DESY, and Valencia.
This work was
partially supported by U.S. Dept. of Energy Contract No. DE-FG 02-
96ER40291.

\end{document}